\documentclass[letterpaper, 10 pt, conference, final, twocolumn]{ieeeconf}  

\usepackage{amsmath}
\usepackage{xcolor}
\usepackage{cite}

\IEEEoverridecommandlockouts                              

\usepackage{graphicx} 
\usepackage{amssymb}  

\usepackage{mathtools,subcaption} 

\newcommand{\setreal}{\mathbb{R}}
\newcommand{\Rnv}{\mathbb{R}^{n_v}}

\newcommand{\vh}{\hat{v}}
\newcommand{\wh}{\hat{w}}
\newcommand{\thetah}{\hat{\theta}}
\newcommand{\col}{\text{col}}

\newcommand{\ion}{{\rm{ion}}}
\newcommand{\syn}{{\rm{syn}}}

\newcommand{\Leak}{{\textrm{leak}}}

\newcommand{\maxcond}{\mu}
\newcommand{\nernst}{E}

\newcommand{\INa}{I_{\rm{Na}}}
\newcommand{\IK}{I_{\rm{K}}}
\newcommand{\IKCa}{I_{\rm{KCa}}}
\newcommand{\ICaT}{I_{\rm{CaT}}}
\newcommand{\ICaL}{I_{\rm{CaL}}}

\newcommand{\gCaL}{\maxcond_{\rm{CaL}}}
\newcommand{\gKCa}{\maxcond_{\rm{KCa}}}

\newcommand{\taumin}{\underline{\tau}}
\newcommand{\taumax}{\overline{\tau}}

\newcommand{\randtau}{p} 
\newcommand{\randsig}{q} 

\newcommand{\obserrrms}{e_{v,\rm{rms}}}

\newcommand{\numredundant}{N}

\title{\LARGE \bf
	Robust online estimation of biophysical neural circuits
}

\author{Raphael Schmetterling$^{1}$, Thiago B. Burghi$^{1}$ and Rodolphe Sepulchre$^{1,2}$
	\thanks{*The research leading to these results has received funding from the European Research Council under the
		Advanced ERC Grant Agreement SpikyControl n.101054323. T.B.B. was supported by the Kavli Foundation.}
	\thanks{$^{1}$Department of Engineering, University of Cambridge, Trumpington Street, Cambridge CB2 1PZ, United Kingdom
		{\tt\small rjzs2@cam.ac.uk} and {\tt\small tbb29@cam.ac.uk}}%
	\thanks{$^{2}$Department of Electrical Engineering, KU Leuven, KasteelPark Arenberg, 10, B-3001 Leuven, Belgium {\tt\small rodolphe.sepulchre@kuleuven.be}} 
}

\begin{document}

	\maketitle
	\thispagestyle{empty}
	\pagestyle{empty}

	\begin{abstract}
		
		The control of neuronal networks, whether biological or neuromorphic, relies on tools for estimating parameters in the presence of model uncertainty. In this work, we explore the robustness of adaptive observers for neuronal estimation. Inspired by biology, we show that decentralization and redundancy help recover the performance of a centralized recursive mean square algorithm in the presence of uncertainty and mismatch on the internal dynamics of the model.

	\end{abstract}

	
	\section{INTRODUCTION}
	With the recent advancements in our ability to record and manipulate neural activity \cite{chen2017neural}, there is a growing call for control and systems tools that can make use of this new technology \cite{tang2018colloquium} for applications including brain-machine interfaces \cite{homer2013sensors} and the treatment of neuronal diseases \cite{ng2016implantable}.
	At the same time the field of neuromorphics, that develops electronics inspired to varying degrees by neuroscience, is seeing rapid growth, with applications in event-based cameras \cite{gallego2020event}, low-power machine learning \cite{tavanaei2019deep}, and many more \cite{schuman2017survey}. Yet, the design of control systems that could interconnect physical neuron-like spiking sensors and actuators through spiking computations is still in its infancy \cite{sepulchre2022spiking}.
	
	The first step in such control tasks is often to obtain model estimates, and there is an extensive literature on fitting neuron models by batch estimation \cite{van2008automated,meliza2014estimating,nogaret2016automatic,abu2019optimal}. A downside of such methods is the fact that they are not able to track time-varying model parameters, which is often necessary to characterize neural behavior.
Our starting point for the present paper is the recent work \cite{burghi2021adaptive}, which proposed an adaptive observer for conductance-based neuron models capable of estimating and tracking model parameters in real time. This method was applied in \cite{schmetterling2022adaptive} to solve classical control problems by adaptively controlling the maximal conductance parameters of these models, an approach aligned with the biological concept of neuromodulation \cite{sepulchre2019control}.

A critical property for practical application of such adaptive methods is to ensure robustness to model uncertainty. Conductance-based models are built from the parallel interconnection of distinct current sources whose specific kinematics and activation range are only approximately known and variable across implementations. The objective of the present paper is to assess the robustness of adaptive estimation against uncertainty in the internal dynamics of conductance-based models. We investigate the effect of uncertainty in a typical neuronal behavior, namely the neuromodulation of a neuron from spiking to bursting by varying slow calcium conductances. 

When the internal dynamics of a neuronal model (also called the channel kinetics) is assumed to be known, \cite{burghi2021adaptive} showed that a simple adaptive observer equivalent to the Recursive Least Squares (RLS) method can be used to estimate the remaining parameters (maximal conductances). Although \cite{burghi2021adaptive} also presents a more elaborate scheme for estimating internal dynamics parameters with local guarantees, here we focus on the basic RLS scheme. In agreement with classical RLS analysis \cite{haykin2002adaptive}, we show empirically that the RLS scheme is sensitive to uncertainty. Our main result is to show that a good estimation performance can be recovered by the combination of two factors that avoid the need for an internal dynamics estimation: first, by decentralizing the RLS scheme as investigated in \cite{burghi2022distributed} with the objective of reducing computational complexity; and second, by introducing redundancy in the estimated model.  
The positive role of redundancy for robustness and adaptation has been extensively demonstrated in neurophysioloogy \cite{drion2015ion,marder2014neuromodulation}. Here, we  mimic biological redundancy by sampling redundant models of ionic currents from a given distribution. This idea can be compared to the method of random features, where random samples of a particular type of basis function are used to solve a regression problem \cite{rahimi2007random}. Our redundant model structure approach can also be related to the ensemble Kalman filter \cite{taghvaei2022optimality}, as well as the more general feedback particle filter \cite{yang2013feedback}. Here, however, the gradient of the observed variable is a function not just of the hidden variables (the internal dynamics), but also of the observed variable itself.
 
    Using the rms (root mean square) observer output error as our performance measure, we study the robustness of the different types of adaptive observer algorithms for conductance-based models in the presence of model error.
    We show that the distributed version of the observer is more robust than the centralized one, and that introducing redundancy to the model structure, according to our proposed approach, further improves this robust behavior. 
	
	\section{BIOPHYSICAL NEURON MODELS}
	\label{sec:biophysical}
	We briefly recall the biophysical conductance-based models of neuron networks. The membrane potential $v_i$ of neuron $i$ in such a network obeys the dynamics
	\begin{equation}
		c_i \, \dot{v}_i = 
		- I_{\Leak,i}
		- \sum_{\ion \in \mathcal{I}} 
		I_{\ion, i}
		-\sum_{\syn \in \mathcal{S}} 
		\sum_{k \neq i} 
		I_{\syn,i,k}
		+
		u_i
	\end{equation}
	where $c_i > 0$ is a capacitance, and each current in the circuit is ohmic in nature. The set $\mathcal{I}$ collects \textit{ionic currents}, while $\mathcal{S}$ collects \textit{synaptic currents}. The \textit{leak current} has a constant conductance and is given by
	\begin{equation*}
		I_{\Leak,i} = \maxcond_{\Leak,i} (v_i - \nernst_{\Leak,i}).
	\end{equation*}
	The ionic and synaptic currents have conductances that are nonlinear and voltage-dependent. The intrinsic ionic currents are modelled by
	\begin{subequations}
		\label{eq:current_cb}
		\begin{align}
			\label{eq:ion_currents}
			I_\ion &= \maxcond_\ion \, m_\ion^{p_\ion} \, h_\ion^{q_\ion} \, (v - \nernst_\ion) \\[.5em]
			\label{eq:activation}
			\tau_{m,\ion}(v) \dot{m}_\ion &= -m_\ion+ \sigma_{m,\ion}(v) \\[.5em]
			\label{eq:inactivation}
			\tau_{h,\ion}(v) \dot{h}_\ion &= -h_\ion+ \sigma_{h,\ion}(v).
		\end{align}
	\end{subequations}
	The constants $\maxcond_{\ion}>0$ 
	and $\nernst_{\ion}\in\setreal$ are called
	(intrinsic) \textit{maximal conductances} and 
	\textit{reversal potentials}, respectively. Note we have dropped the index $i$ to simplify the notation. 
	
	The static \textit{activation functions}
	$\sigma_{m,\ion}(v)$
	and 
	$\sigma_{h,\ion}(v)$,
	and
	\textit{time-constant functions} $\tau_{m,\ion}(v)$ 
	and $\tau_{h,\ion}(v)$, model the nonlinear gating 
	of the ionic conductance. Because 
	$\sigma_{m,\ion}:\setreal\to(0,1)$ 
	and 
	$\sigma_{h,\ion}:\setreal\to(0,1)$ 
	are monotonically increasing and decreasing, respectively, the states $m_\ion$ and $h_\ion$ are called \textit{activation} and  \textit{inactivation gating variables}, respectively. 
	The time-constant functions vary in shape, but always respect the bounds
	\begin{equation*}
		0 < \taumin_\ion \le \tau_{m,\ion}(v),
		\tau_{h,\ion}(v) \le \taumax_\ion
	\end{equation*}
	for all $v \in \setreal$ and some $\taumin_\ion,\taumax_\ion > 0$. 
	The exponents $p_\ion$ and $q_\ion$ in \eqref{eq:ion_currents} are
	natural numbers (including zero). Each gating variable can be thought of as ``opening'' or ``closing'' a particular ionic channel; exponents greater than one represent multiple identical ``gates'' in series \cite{dayan2005theoretical}. In this paper the exponents always take value unity, which simplifies the notation without losing any behaviours of interest.
	
	We demonstrate our results on a neuronal model  that includes five typical ionic currents of a bursting neuron: a
	transient sodium current $\INa$,
	a potassium current $\IK$, a T-type calcium current $\ICaT$, an L-type calcium current $\ICaL$ and a calcium-activated potassium current $\IKCa$. We therefore have
	$\mathcal{I} = \{\rm{Na},\rm{K},\rm{CaT},\rm{CaL},\rm{KCa} \}$. 
	The voltage dynamics of a single, isolated
	neuron (no synaptic currents) are given by 
	\begin{align*}
			c \, \dot{v} = 	
			&-\maxcond_{\rm{Na}} m_{\rm{Na}} h_{\rm{Na}} 
			(v-\nernst_{\rm{Na}}) \\
			&-\maxcond_{\rm{K}} m_{\rm{K}}
			(v-\nernst_{\rm{K}}) \\
			&-\maxcond_{\rm{CaT}} m_{\rm{CaT}} h_{\rm{CaT}} 
			(v-\nernst_{\rm{Ca}}) \\
			&-\maxcond_{\rm{CaL}} m_{\rm{CaL}}
			(v-\nernst_{\rm{Ca}}) \\
			&-\maxcond_{\rm{KCa}} \sigma_{\rm{KCa}}([Ca]) 
			(v-\nernst_{\rm{K}})
			-\maxcond_\Leak(v-\nernst_\Leak)
			+ u,
	\end{align*}
	where $[Ca]$ is the calcium concentration, governed by
	\begin{align*}
		\tau_{\rm{Ca}} \dot{[Ca]} = &-0.03 m_{\rm{CaT}} h_{\rm{CaT}} 
		(v-\nernst_{\rm{Ca}}) \\&-0.3 m_{\rm{CaL}}
		(v-\nernst_{\rm{Ca}}) - [Ca],
	\end{align*}
	with $\tau_{\rm{Ca}}$ a constant. For a full list of parameters used, we refer the reader to the Julia code attached to this paper.\footnote{https://github.com/RJZS/robust-neuron-estimation}
	
\begin{figure}
		\centering
		\includegraphics[width=3.35in]{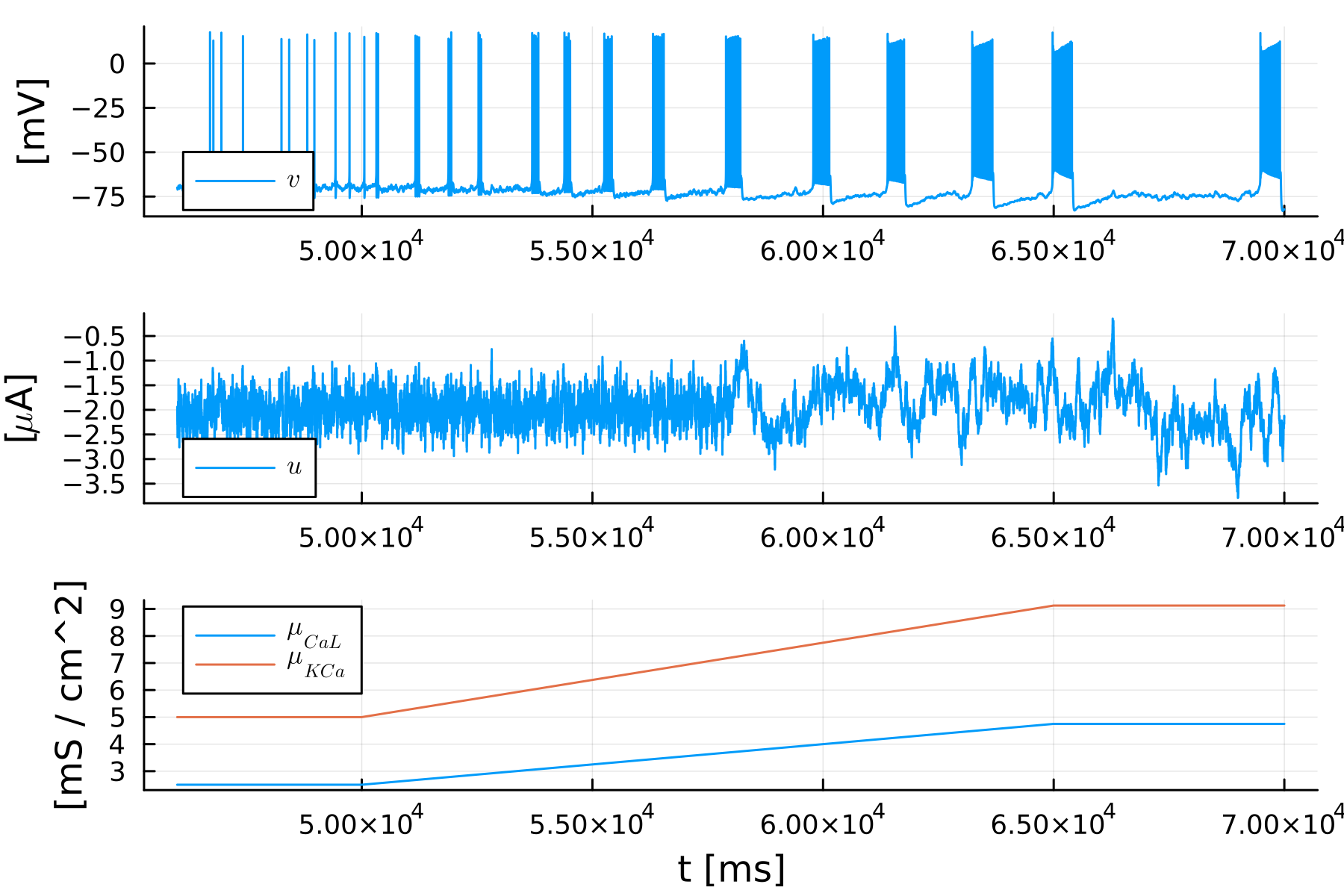}
		\caption{Fixed scenario for the illustrations of this paper. Top: membrane potential. Middle: input current. Bottom: maximum conductance of L-type calcium and calcium-activated potassium  (the other parameters remain constant).   Note that the input's fluctuations change during the simulation to reflect the difference between spike and burst excitability. }
		\label{fig:task}
	\end{figure}

	We consider a fixed scenario illustrated in Fig. \ref{fig:task}: the neuron is driven by a known fluctuating input calibrated to expose its excitable behavior. 
	The maximal conductance of L-type calcium, $\gCaL$, and calcium-activated potassium, $\gKCa$, are ramped up during the simulation, which results in a modulation from spike excitability to burst excitability. Such neuromodulation is a key cellular mechanism in neurophysiology \cite{drion2018switchable,drion2019cellular}.
	
	\section{ADAPTIVE RECURSIVE LEAST SQUARES ESTIMATION}
	
	We briefly summarize the simple centralized RLS-based observer of \cite{burghi2021adaptive}. The starting observation is that conductance-based neuron networks obey the following non-linear state-space form:
	\begin{align}
		\dot{v} &= \Phi^T(v, w, u) \theta + a(v, w, u) \label{eq:phitheta} \\
		\dot{w} &= g(v, w) \label{eq:internal}
	\end{align}
	where $v \in \Rnv$ is a state vector and the system output, representing the neuronal membrane voltages; $w$ is an internal dynamics state vector, collecting the dynamics of all gating variables and of calcium concentration; and $u \in \Rnv$ is a control input vector. We also have a vector $\theta$ collecting all the maximal conductances $\mu_\ion$ of the system.  For simplicity we will study the single-neuron case $n_v~=~1$, but the results generalize easily to networks of arbitrary size. For the example neuron in Section \ref{sec:biophysical}, we have\[\theta = \text{col}\big( \mu_{\rm Na},\mu_{\rm K}, \dotsc, \mu_{\rm leak} \big)\] and
 \[ w = \text{col}\big( m_{\rm Na},h_{\rm Na}, m_{\rm K}, \dotsc, [Ca] \big).\]

We use an adaptive observer to estimate the linear parameters $\theta$, in our case the vector of maximal conductances $\mu_\ion$. We choose maximal conductances for the unknown parameters because the modulation of conductance properties is a key control mechanism of neurophysiology, where it is performed by neuromodulators such as dopamine and serotonin \cite{eppinger2011neuromodulation}. Neuromodulation is a well-studied phenomenon; it is essential to the function of all nervous systems \cite{marder2012neuromodulation}.
	
	The adaptive observer estimates the system's state and parameters, given measurements of the input (current) $u(t)$ and output (voltage) $v(t)$.
	It relies on the assumptions that the system trajectories evolve in a compact positively invariant set, and that the internal dynamics (\ref{eq:internal}) are exponentially contracting, uniformly in $v$.
	
	The simple RLS-based centralized observer \cite{burghi2021adaptive} takes the following form:
	\begin{align}
		\dot{\vh} &= \Phi^T(v,\wh,u)\thetah + a(v,\wh,u) + 			\gamma(I+\Psi^T P \Psi)(v-\vh) \\
		\dot{\wh} &= g(v,\wh) \label{eq:internal_obs} \\
		\dot{\thetah} &= \gamma P \Psi (v-\vh)
	\end{align}
	where $\gamma > 0$ is a constant gain, and the matrices $P$ and $\Psi$ evolve according to
	\begin{align*}
		\dot{\Psi} &= -\gamma\Psi + \Phi(v,\wh,u) \\
		\dot{P} &= \alpha P - \gamma P \Psi^T \Psi P \;\;\;\;\;\;P(0) \succ 0
	\end{align*}
	with $\gamma > \alpha > 0$. Notice that $\Psi$ is a low-pass filtered version of $\Phi$, while $P$ can be interpreted as a running estimate (with a forgetting factor) of the parameter covariance matrix \cite[Chapter~2]{aastrom2013adaptive}. Note also the use of output injection, that is the injection of true $v$ into the $\hat{v}$ and $\hat{w}$ dynamics.
	With the assumptions listed above, and a standard persistent excitation condition, it can be shown that the adaptive observer state vector $\col(\vh(t),\wh(t),\thetah(t))$ converges to $\col(v(t),w(t),\theta(t))$ exponentially fast as $t \to \infty$.
	See \cite{burghi2021adaptive} for a contraction-based proof of convergence.
	
\begin{figure}
		\centering
		\includegraphics[width=3.35in]{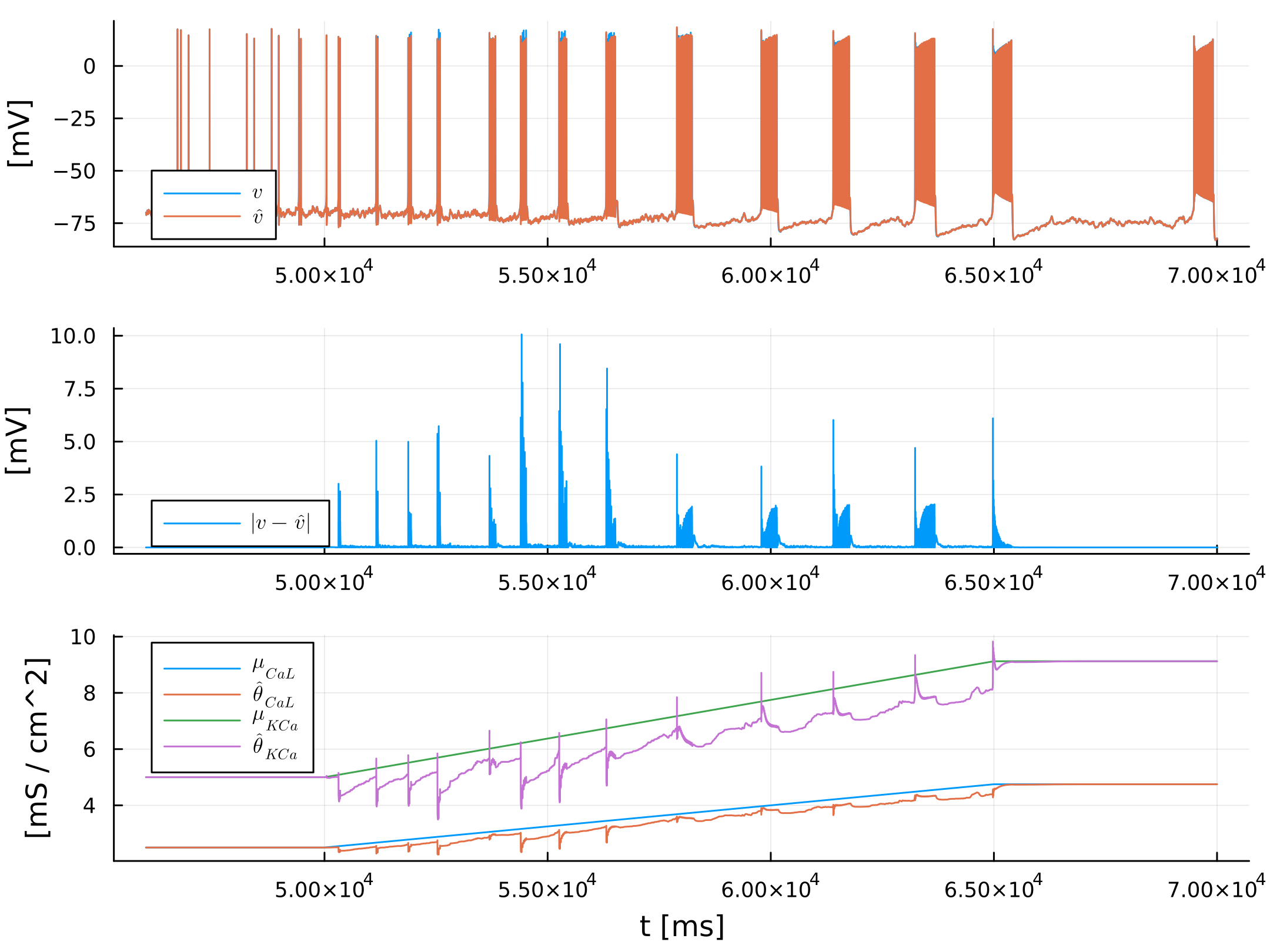}
		\caption{RLS estimation in the absence of model error. Parameter and output estimates converge to their true values, and remain there barring transient errors while $\gCaL$ and $\gKCa$ are modulated. Top: true voltage $v$ and its estimate $\hat{v}$. Middle: absolute observer error $|v-\hat{v}|$. Bottom: L-type maximal conductance $\gCaL$, its estimate $\hat{\theta}_{\rm{CaL}}$, calcium-activated potassium maximal conductance $\gKCa$, and its estimate $\hat{\theta}_{\rm{KCa}}$.}
		\label{fig:noerr}
	\end{figure}

	Fig. \ref{fig:noerr} shows that the observer is able to learn the task of Fig. \ref{fig:task}, in the absence of model error. Errors in the voltage and parameter estimates are only present transiently, while the maximal conductances vary. For brevity, the figure shows only the two parameters that are modulated, but recall all the maximal conductances are estimated.
	
	\section{ROBUSTNESS OF THE OBSERVER}
	
	To investigate the robustness of the centralized observer against variability  in the internal dynamics, we introduce the random variables $\randtau \sim U(1-r_,1+r)$ and $\randsig \sim U(-s, s)$, where $U(a,b)$ represents the uniform distribution with support $[a,b]$. We define stochastic versions of the gating variable dynamics (\ref{eq:activation})-(\ref{eq:inactivation}) as follows:
		\begin{subequations}
		\begin{align}
			\label{eq:randactivation}
			\randtau_{m,\ion} \, \tau_{m,\ion}(v) \dot{m}_\ion &= -m_\ion+ \sigma_{m,\ion}(v-\randsig_{m,\ion}) \\[.5em]
			\label{eq:randinactivation}
			\randtau_{h,\ion} \, \tau_{h,\ion}(v) \dot{h}_\ion &= -h_\ion+ \sigma_{h,\ion}(v-\randsig_{h,\ion}).
		\end{align}
	\end{subequations}
	The effect of the random variables is to respectively scale and shift the time-constant and activation functions. We collect the samples into vectors $p, q \in \setreal^{n_w}$ and the randomized dynamics into a function $g(v,\hat{w}; p, q)$. We replace the observer's internal dynamics (\ref{eq:internal_obs}) with
    \begin{equation}
        \dot{\hat{w}} = g(v,\hat{w}; p, q). \label{eq:new_g}
    \end{equation}
    
    This introduces mismatch between the true dynamics (\ref{eq:internal}), which remain deterministic, and the observer's model of these dynamics. In this paper, we take $r = 0.04$ and $s = 4 mV$.
	
\begin{figure}
		\centering
		\includegraphics[width=3.35in]{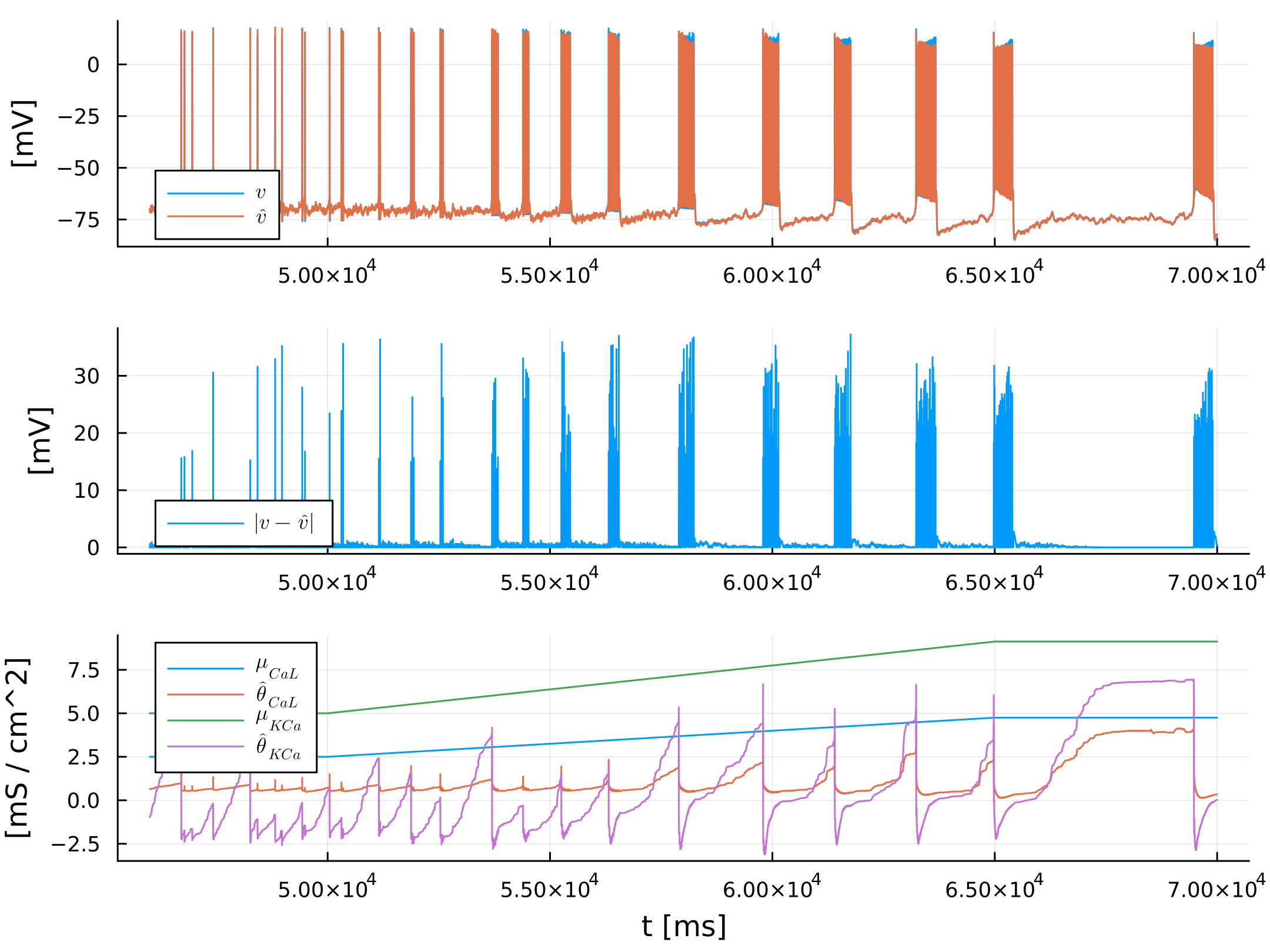}
		\caption{The centralized observer is fragile to model uncertainty. Top: true voltage and its estimate. Middle: absolute observer error. Bottom: time-varying maximal conductances and their estimates.}
		\label{fig:centr}
	\end{figure}

	Fig. \ref{fig:centr} illustrates the observer's performance with one set of samples for model error. We chose $\gamma=8$ and $\alpha=0.005$. Higher $\gamma$ reduce observer error but are sensitive to noise measurement, hence we use the same value throughout the paper for a fair comparison. The value for $\alpha$ was tuned manually to optimize performance. 
 
    We take our performance measure to be $\obserrrms$, the rms value of the observer error over the duration of the simulation in  Fig. \ref{fig:task}. This is computed as $(\sum_{t=1}^T (v(t) - \hat{v}(t))^2/N)^{\frac{1}{2}}$ where $N$ is the number of simulation time steps; the step size is $\Delta t = 0.1$. Although output injection ensures that the spike and burst estimates align, there is significant error in the output and parameter estimates. The parameter estimates fail to track the modulation of calcium currents, limiting any practical use of the online observer.
	

	\section{DISTRIBUTED OBSERVER} \label{distributed}
	
	The centralized observer introduced above is based on the recursive least squares algorithm \cite{burghi2021adaptive}. In the presence of zero-mean, independent, identically distributed additive noise, the matrix $P$ is therefore proportional to the covariance matrix of the empirical estimate of $\theta$ \cite[Chapter~2]{aastrom2013adaptive}. When we introduce model mismatch, this second-order information becomes unreliable. A first remedy to increase robustness is to decentralize the parameter estimation.  The core idea is to approximate the matrix $\Psi \Psi^T$, which appears in the update equation for $P$, by its (potentially block) diagonal elements to yield a decentralized learning rule. We thus replace the $n_\theta \times n_\theta$ matrix $P$ with $n_\theta$ scalars $P_i$. This idea was already explored in \cite{burghi2022distributed} with the goal of reducing the algorithmic complexity of the estimator from O$(n_\theta^2)$ to O$(n_\theta)$, a significant benefit for more complex neurons and networks with many synapses.
	
	The distributed observer with model mismatch has the form:
	\newline
	\begin{align}
		\dot{\vh} &= \sum_{j=1}^{n_\theta} \Phi^T_j(v,\wh^j,u)\thetah_j + a(v,\wh,u) \label{eq:distobs_1} \\
		& \hspace{4em} +(\gamma_0 I+\sum_{j=1}^{n_\theta} \gamma_j \Psi^T_j P_j \Psi_j)(v-\vh) \nonumber \\
		\dot{\wh}^j &= g_j(v,\wh^j; p_j, q_j) \label{eq:distobs_2} \\
		\dot{\thetah}_j &= \gamma_j P_j \Psi_j (v-\vh) \label{eq:distobs_3}
	\end{align}
	where $\gamma_0, \gamma_1, \ldots, \gamma_{n_\theta} > 0$ are constant gains. The matrices $P_j$ and $\Psi_j$ evolve according to
	\begin{align*}
		\dot{\Psi}_j &= -\gamma_j\Psi_j + \Phi_j(v,\wh^j,u) \\
		\dot{P}_j &= \alpha_j P_j - \alpha_j P_j \Psi^T_j \Psi_j P_j \;\;\;\;\;\;P_j(0) \succ 0
	\end{align*}
	with $\alpha_j > 0 \; \forall j$.

    For the example neuron in Section \ref{sec:biophysical}, for instance, we have \[\theta_1 = \mu_{\rm Na}, \; \theta_2 = \mu_{\rm K}, \, \dotsc , \, \theta_{n_\theta} = \mu_\Leak\] and 
    \[w^1 = \text{col}(m_{\rm Na},h_{\rm Na}), w^2 = m_K, \dotsc, w^{n_{\theta}} = \emptyset \]
    
\begin{figure}
		\centering
		\includegraphics[width=3.35in]{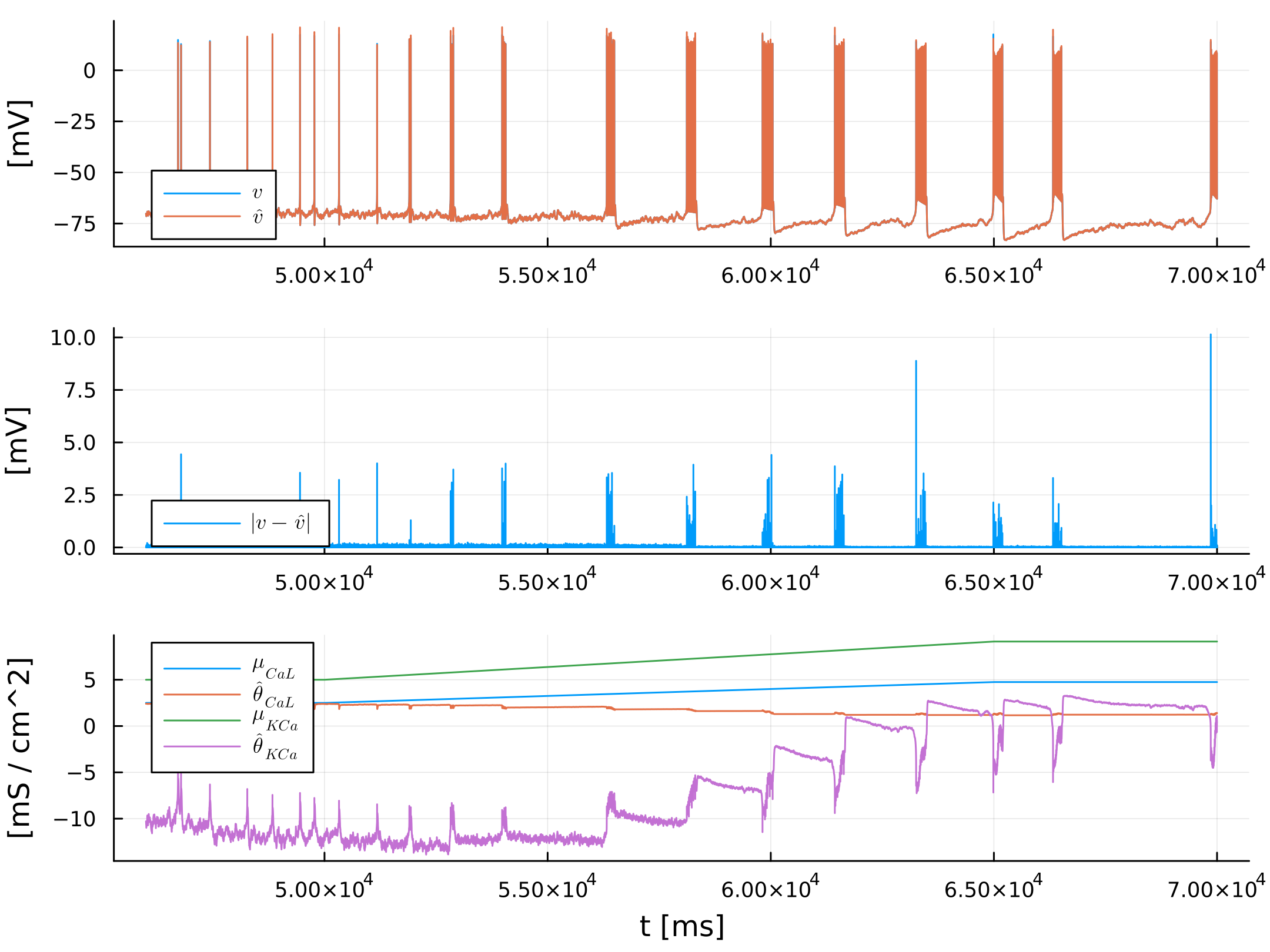}
		\caption{Distributed parameter estimation in the scenario of Fig. \ref{fig:task}. Top: true voltage and its estimate. Middle: absolute observer error. Bottom: $\gCaL$ and $\gKCa$ and their estimates. Although there is some observer error, it is significantly smaller as reflected in the $v$ and $\hat{v}$ spikes being almost indistinguishable. }
		\label{fig:distr}
	\end{figure}

	Fig. \ref{fig:distr} shows a representative example of the distributed observer's performance in the presence of randomly-sampled model error. We set $\gamma_j=8$ for all $j$, the same as with the centralized observer for a fair  comparison. We set by hand tuning $\alpha_j=2\times 10^{-4}$ for all $j$. To compare the performance of the two observers, we compute the mean value of $\obserrrms$ across twenty trials.
	The results are shown in the first two columns of Table \ref{table:results_full}. The observer error is indeed reduced by the use of the distributed observer. The parameter estimates have also improved, as they oscillate less and are therefore more meaningful. We do not necessarily expect the parameter estimates to settle near the true values, as the observer error can be reduced by exploiting the biological redundancy between currents. In the next section, we exploit the theme of redundancy to further improve robustness.

	\section{REDUNDANCY}
	
	In biology, redundancy between ionic currents allows them to compensate for each other to achieve desired behavioral properties \cite{drion2015ion}. Redundancy is known to play a crucial role in the homeostasis of neuronal function in spite of the highly variable ion channel density, both across time and from animal to animal \cite{marder2014neuromodulation}.

    Redundancy can also be exploited to increase the robustness of estimation to model uncertainty. To test that idea, we consider an augmented observer model that includes for each gating variable $\numredundant$ equations of the form (\ref{eq:randactivation}) or (\ref{eq:randinactivation}), each with its own samples for $p$ and $q$. That is to say, each element of (\ref{eq:distobs_2}) is replaced with $\numredundant$ elements 
    \begin{equation}
        \dot{\hat{w}}^{j,i} = g_j(v,\hat{w}^{j,i};p_i^j,q_i^j)
    \end{equation}
    for $i = \{1,\ldots,\numredundant\}$.
    We can consider this step as replacing a single estimate of the gating variable $m_\ion$ or $h_\ion$ with $\numredundant$ particles of the same, and the resulting algorithm analogous to an ensemble observer. Note that we do not change the model of the neuron being observed.

    Every ionic current now has $\numredundant$ corresponding terms in the $\vh$ dynamics. Hence, we replace (\ref{eq:distobs_1}) with
    \begin{align}
    \nonumber
        \dot{\vh} = &\sum_{j=1}^{m} \bar{\Phi}_{j}^T(v,\wh^{j},u)\thetah_{j} 
        + a(v,\wh,u)  \\
        & \hspace{4em} +(\gamma_0 I+\sum_{j=1}^{m} \gamma_{j} \bar{\Psi}^T_{j} \bar{P}_{j} \bar{\Psi}_{j} 
        )
        (v-\vh), 
    \end{align}
    where $j \in \mathcal{I}$ is the set of membrane currents. We now have 
    \[
    \bar{\Phi}_j = \text{col}\big(\Phi_{j}(v,\wh^{j,1},u),\ldots,\Phi_{j}(v,\wh^{j,N},u)\big) \in \setreal^{\numredundant}
    \]
    for $j=1,\dotsc,n_\theta-1$, and 
    \[\bar{\Phi}_{n_\theta}(v,\hat{w}^{n_\theta},u) = \bar{\Phi}_{n_\theta}(v) = -(v-E_\Leak)\] with the latter corresponding to the leak current regressor. $\bar{P}_j$ is now an $\numredundant \times \numredundant$ diagonal matrix with leading diagonal $\text{col} \big (\bar{P}_j^1,\ldots,\bar{P}_j^\numredundant \big )$.
    
    We also have 
    \[
    \thetah_j = [\thetah_j^1,\dotsc,\thetah_j^N],
    \]
    the set of parameter estimates corresponding to a particular maximal conductance of the reference neuron. We define the empirical mean over this set,
    \begin{equation*}
        \bar \thetah_j = \frac{1}{\numredundant}\sum_{i=1}^\numredundant \thetah_{j}^i,
    \end{equation*}
    which provides an estimate of the scaled maximal conductance $\mu_j / \numredundant$.
 
	We will apply redundancy only to the distributed observer. The diagonal nature of the distributed observer makes it scalable with respect to the increased number of states and parameters.
	
	Redundancy is of course antagonist to persistency of excitation, and indeed a naive implementation of the observer leads to situations where some estimated maximal conductances become negative, causing instability. 
	
	To prevent the divergence of a redundant estimator, we modify the $\thetah_j$ update law (\ref{eq:distobs_3}) to include a consensus term, that regularizes the variance of the redundant parameters. The update law for the $i$th redundant element of $\thetah_j$ is now
	\begin{equation}
		\dot{\hat{\theta}}_{j}^i = \gamma_j \bar P_{j}^i \bar \Psi_{j}^i (v - \hat{v}) - \beta (\hat{\theta}_{j}^i - \bar \thetah_j).
	\end{equation}
	
\begin{figure}
		\centering
		\includegraphics[width=3.35in,height=3.35in]{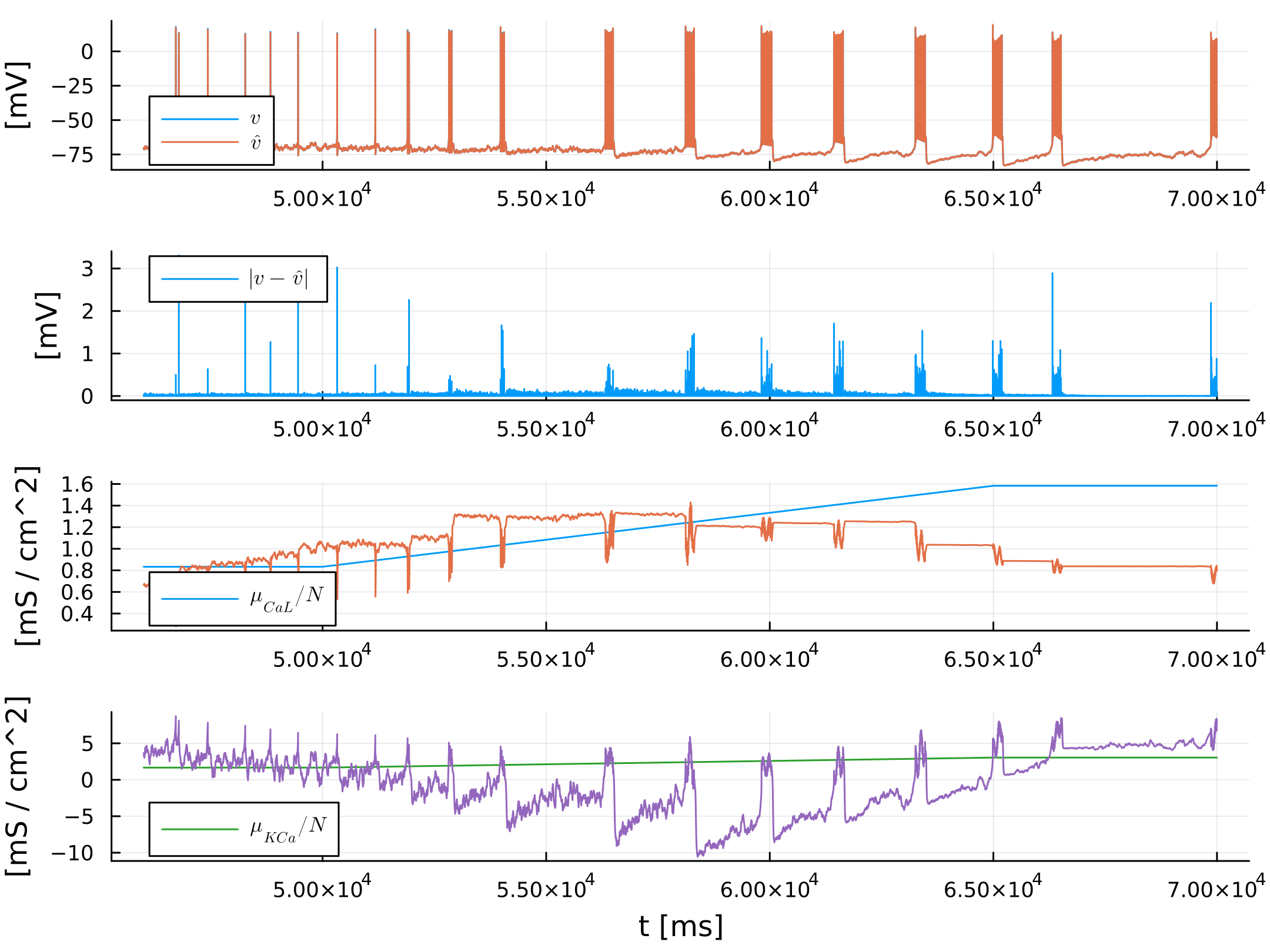}
		\caption{Distributed and redundant online estimation in the scenario of Fig. \ref{fig:task}, in the presence of model error and with $\numredundant = 3$. Top: true voltage and its estimate. Upper middle: absolute observer error. Lower middle: the empirical mean estimate $\bar \thetah_{\rm{CaL}}$; for comparison, the scaled true parameter $\gCaL/\numredundant$ is also shown. We plot the \textit{scaled} parameter to emphasise that we have replaced each conductance with $N$ separate conductances. Bottom: the same for $\gKCa$. Labels for the parameter estimates are omitted.}
		\label{fig:distredundant}
	\end{figure}

	Fig. \ref{fig:distredundant} illustrates the performance of the redundant estimator in the presence of model error. We set $\gamma_j$ and $\alpha_j$ as in section \ref{distributed}. We chose by hand tuning $\beta = 5\times 10^{-5}$, a value low enough that the redundant parameters take on distinct values.
 
	The mean rms error across twenty trials is provided in Table \ref{table:results_full}. The third and forth columns refer to the final algorithm, respectively with 3 and 9 redundant elements per gating variable. As expected, we see significant improvement of the observer error when redundancy is introduced, and a greater improvement with more redundancy. 
 
 We see however in Fig. \ref{fig:distredundant} that the redundant terms do not necessarily directly track modulation. This is as the observer is exploiting redundancy across all terms to minimise voltage error, not just those corresponding to a single ionic current.

	
	\begin{table}
		\renewcommand{\arraystretch}{1.3}
		\caption{Comparison of all three observers, showing the mean and standard deviation of the rms voltage error $\obserrrms$ (in $mV$) across twenty trials.}
		\label{table:results_full}
		\centering
		\begin{tabular}{|c|c|c|c|c|} \hline
			 & Centralized & Distributed & $\numredundant=3$ &  $\numredundant=9$\\ 
			\hline\hline
			Mean & 1.15 & 0.0788 & 0.0280 & 0.0241  \\ \hline
            Standard Deviation & 0.14 & 0.017 & 0.0056 & 0.0030  \\ 
            \hline
		\end{tabular}
	\end{table}

	\section{DISCUSSION}
	
	Our results suggest that an adaptive observer can be used to estimate neuronal parameters, even in the presence of modelling error. The chosen model of uncertainty is plausible in a neuromorphic context where a key cause of error is transistor mismatch introduced during manufacturing \cite{serrano1999systematic}. This component imprecision is one of the main challenges facing designers of neuromorphic hardware \cite{liu2010neuromorphic}.
	In a biological context, the cell is of course part of a living system and is therefore time-varying. The online nature of the observer ensures that our results extend to this time-varying case.
	
	In future work, the results above should also be tested against measurement error.  Noisy voltage measurements  introduce trade-offs  in the design parameters. For a rigorous examination of the impact of noise on the system identification of conductance-based models, we refer the reader to \cite{burghi2021feedback}.
	
	The performance metric of our study was the rms observer. This is a reasonable first step to make a quantitative comparison of different observers. However, it is only a proxy of the practical objective to estimate parameters in order to track neuromodulation in an experimental setup  or to use learning experiments for hardware implementations of neuromorphic neurons.  This will be the topic of future research.
	
	
	
	
	\addtolength{\textheight}{-2cm} 
	


	
	
	

	\bibliographystyle{IEEEtran}
	\bibliography{IEEEabrv,IEEEexample,cdc_refs}

\end{document}